\documentclass[10pt,conference]{IEEEtran}
\IEEEoverridecommandlockouts
\pdfoutput=1
\usepackage{cite}
\usepackage{amsmath,amssymb,amsfonts}
\usepackage{algorithmic}
\usepackage{graphicx}
\usepackage{textcomp}
\usepackage{xcolor}
\usepackage{hyperref}
\usepackage{booktabs}

\usepackage{booktabs} 
\usepackage{array}
\usepackage{subcaption}
\usepackage{cleveref}
\usepackage[colorinlistoftodos]{todonotes}
\usepackage[justification=centering]{caption}

\usepackage{balance}

\usepackage{listings}
\begin{document}

\title{Executability of Python Snippets in Stack Overflow
}


\author{\IEEEauthorblockN{Md Monir Hossain\IEEEauthorrefmark{1}, 
Nima Mahmoudi\IEEEauthorrefmark{1},
Changyuan Lin\IEEEauthorrefmark{1}, 
Hamzeh Khazaei\IEEEauthorrefmark{1}
and Abram Hindle\IEEEauthorrefmark{3}}
\IEEEauthorblockA{\IEEEauthorrefmark{1}\textit{Department of Electrical and Computer Engineering}}
\IEEEauthorblockA{\IEEEauthorrefmark{3}\textit{Department of Computing Science}}
\textit{University of Alberta}, Edmonton, Canada \\
Email:\{mdmonir,nmahmoud,changyua,hkhazaei,abram.hindle\}@ualberta.ca}


\maketitle

\begin{abstract}
Online resources today contain an abundant amount of code snippets for documentation, collaboration,
learning, and problem-solving purposes. Their executability in a `plug and play' manner enables us to confirm their quality and use them directly in
projects. But, in practice that is often not the case due to several requirements violations or incompleteness. However, it is a difficult task
to investigate the executability on a large scale due to different possible errors during the execution. We have developed a scalable framework to
investigate this for SOTorrent Python snippets. We found that with minor adjustments, 27.92\% of snippets are executable. The executability hasn't
changed significantly over time. The code snippets referenced in GitHub are more likely to be directly executable. But executability doesn't affect the chances of the answer to be
selected as the accepted answer significantly. These properties help us
understand and improve the interaction of users with online resources that include code snippets.
\end{abstract}


\section{Introduction}
In code development, it is a common practice to learn and develop using code samples from online sources like software repositories, documentations, Question Answering forums, version control services, etc~\cite{gharehyazie2017some,abdalkareem2017code}. Question Answering Websites have risen to prominence in recent years along with the increase in people's online presence and open-source software systems.
The construction and operation of question answering forums have changed from one-off isolated interactions to continuous community-driven interactions, acting like a collaborative knowledge sharing hub. This has created an interactive environment where both questions and answers can evolve and improve over time in the community.

Stack Overflow, the flagship of Stack Exchange, is one of the most popular question answering forums in the programming domain. It employs the means of reputation, upvote, downvote, acceptance, bounty, and moderation to make an interactive experience. The questions and answers both contain code snippets along with contextual texts. The comments also impact the questions and answers by asking for explanations or providing argument. 

Stack Exchange network provides a collection of user contributed contents' dump online which includes Stack Overflow. SOTorrent~\cite{baltes2018sotorrent} was developed based on that to answer questions on the evolution of Stack Overflow Posts. It captures version change of posts, texts and codeblocks along with external URLs. The research questions we are interested to answer using this dataset are:


\textbf{RQ1: What percentage of SO Python snippets are executable directly with minor adjustments? What are the reasons that they are not?} The motivation here is to understand how well and complete the written code snippets are and what the common things are that users miss out which could have made the codes directly executable. Knowing the degree of adjustment they require in making them executable, gives us an idea of the effort the task demands.

\textbf{RQ2: Can we find a trend which reflects the change in the executability of python snippets over time and make conclusions from that?} The motivation is to find out if the community has adopted a more structured approach in code sharing over time. Communities like GitHub may value direct executabilty more than general users who should favor textual explanation more.  

\textbf{RQ3: Can we find a relationship between the executability of code snippets and GitHub reference?} The motivation here is to find if referring to GitHub indicates a higher probability of directly executable code snippet. We believe that code blocks that need fewer modifications to execute successfully might be more likely to be used in GitHub projects.

\section{Related Work}

Although the reuse of code snippets introduces complications like licensing~\cite{ragkhitwetsagul2018awareness,an2017stack,baltes2017attribution,ragkhitwetsagul2018toxic}, maintainability, reliability~\cite{zhang2018code}, bugs, or being outdated, their positive impact in developer community is evident. And this impact can be classified in categories like `learn as you go', `clarification and extension' or `remember things you forget'~\cite{brandt2009two}. Due to the effectiveness of code snippets, several works have tried to mine these from various sources to answer questions pertinent to software engineering. These questions deal with aspects like code snippet collection, best use, evolution, maintainability, legal implication, etc. Subramanian et al. did an analysis of Stack Overflow code snippets to identify structural relationships from snippets on  questions  relevant to android~\cite{subramanian2013making}. XSnippet was proposed as a context-aware framework to facilitate developers in querying relevant code snippet from a sample repository~\cite{sahavechaphan2006xsnippet}. Keivanloo et al. proposed a method to find out working code examples using code clone detection model that can be used by internet scale source code search engine~\cite{keivanloo2014spotting}. Gistable empirically analyzed the executability of python code snippets from GitHub gist system which found that majority of the python snippets there could be run with minor adjustment and provided the dataset online~\cite{horton2018gistable}. Nasehi et al. did a quantitative analysis on Stack Overflow questions and answers to find out what makes a good code example~\cite{nasehi2012makes}. They found that developers' learning experience can possibly be enhanced by adjusting the attributes of code example and explanation as they relate to the properties of recognized successful answers.

\section{Methodology}

\subsection{Dataset}\label{AA}
The SOTorrent Dataset is available in the form of CSV and XML dumps in Zenodo\footnote{\href{https://zenodo.org/record/1295405\#.XFiEflxKguU}{https://zenodo.org/record/1295405\#.XFiEflxKguU} last accessed: Feb-4-2019.}. Another version of it is available in Google BigQuery
\footnote{\href{https://bigquery.cloud.google.com/dataset/sotorrent-org:2018_12_09}{https://bigquery.cloud.google.com/dataset/sotorrent-org:2018\_12\_09} last accessed: Feb-4-2019}
as public dataset. We have used the `Posts' and `PostBlockVersion' tables from the SOTorrent Dataset (Dated 2018-12-02) in Zenodo for the analysis.  

\subsection{Data Preprocessing}\label{AA}
We extracted the code blocks in the answers with only `python' tag. For this, we first mapped the tags from the questions to associated answers in the `Posts' table and filtered by `python' tag. 

\begin{lstlisting}
CREATE TABLE
  PostsAnswerPython
SELECT
  table1.Id, 
  table1.ParentId, 
  table1.PostTypeId, 
  table1.Score, 
  table2.Tags
FROM
  Posts AS table1
INNER JOIN
  Posts AS table2
ON
  table1.ParentId = table2.Id
WHERE
  table2.Tags LIKE `<python>';
\end{lstlisting}


Then, we merged newly created `PostsAnswerPython' and original `PostBlockVersion' based on PostId. As `PostBlockVersion' contains different `PostBlockTypeId' for texts and codeblocks, we filtered based on that.  

\begin{lstlisting}
CREATE TABLE
  SnippetsAnswerPython
SELECT
  table3.Id, 
  table3.PostId, 
  table3.PredPostBlockVersionId,
  table3.RootPostBlockVersionId, 
  table3.Length, 
  table3.LineCount, 
  table3.PostBlockTypeId,
  table4.Tags, 
  table4.Score, 
  table4.ParentId, 
  table3.Content
FROM
  PostBlockVersion AS table3
INNER JOIN
  PostsAnswerPython AS table4
ON
  table3.PostId = table4.Id
WHERE
  table3.PostBlockTypeId=2;
\end{lstlisting}


After filtering we ended up with 269784 Python snippets on which we performed our analysis. 
\subsection{System Overview}

\begin{figure}[htbp]
\centerline{\includegraphics[width=0.9\columnwidth]{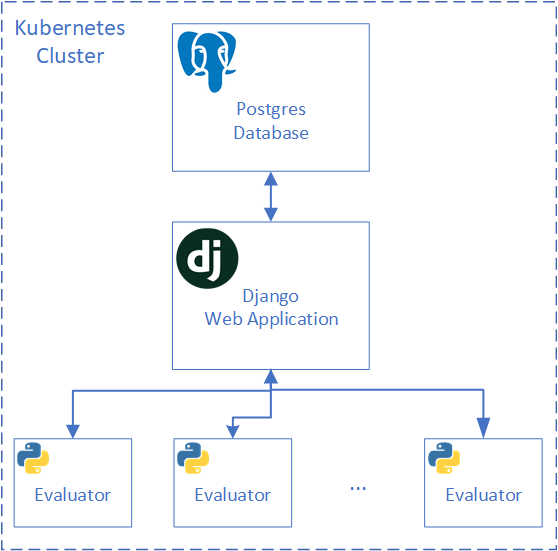}}
\caption{An overview of the developed system architecture.}
\label{fig:overview}
\end{figure}

In this section, we introduce the overview of the scalable platform designed for evaluating the executability of the SOTorrent dataset.
We designed and developed an evaluator for the code snippets that works on docker.
The overview of the designed architecture can be found in \Cref{fig:overview}.
To have a scalable platform, we designed a web application that serves the code snippets to the evaluator instances over a  JSON API over HTTP 
in a Kubernetes cluster. The evaluator instances fetch snippets from the database, evaluate them and then send back the result to the
web application. After the whole dataset is processed, we fetch the results back from the database for further analysis.
Using this architecture, we were able to evaluate the executability of around 50 code snippets per second using 9 VMs each
having 4 cores of vCPU and 15GB of memory. For the evaluation platform, we deployed 12 instances of Django 2.1 as the web 
application, 1 instance of PostgreSQL 9.6.6 as the database and 30 instances of our snippet evaluator.

\subsection{Code Snippet Evaluator}
In this section, we discuss the implementation of the evaluator. As the name suggests, the evaluator is where the executability of code snippets is evaluated. It is also responsible for snippet retrieval, code parsing, snippet execution, result collection, error extraction, exception handling, module installation and posting the evaluation results. We created a docker container which contains Python 3.7, Python 2.7 and pip 19. We analyzed the whole dataset and extracted the top 40 popular Python modules by parsing the statement with import and counting word frequency. In order to make evaluation faster, we installed these 40 popular Python modules in the container in advance.

The evaluator first retrieves a snippet from the API. Then, the snippet is decoded and parsed.
The decoded and parsed snippet is written to a python file. In the next step, the evaluator creates a new process to execute this file. A timeout threshold of 10 seconds is used after which the snippet is considered to be timed out and is not evaluated furthermore. 

After execution, the evaluator collects the results and the return code and sends it back to the web server. We followed the Python errors and built-in exceptions specified in the official documentation of Python 3.7 and Python 2.7 to define 58 status codes, such as `SyntaxError' and `ExitCodeException'. The return code 0 stands for successful execution without any errors. 

In case of an `ImportError' or `ModuleNotFoundError', the evaluator extracts the name of the imported modes and tries to install those
modules by calling pip. The module is then evaluated again
to check if its status has changed. In case the error still persists, we ignore the package installation and send back the results to the server. This process has been dockerized to make it reproducible and easily scalable for larger datasets.
The code used for evaluation is available online\footnote{\href{https://github.com/DDSystemLab/python-executability-analysis}{https://github.com/DDSystemLab/python-executability-analysis} last accessed: Jul-4-2019.}.

\section{Results}

\textit{RQ1: What percentage of SO Python snippets are executable directly with minor adjustments? What are the reasons that they are not?}

We found that after installing the top 40 packages, for Python 2 environment 26\% and for Python 3 environment 23\% of codes snippets were executed without any errors. \Cref{tab:executability-results} depicts a detailed account of Success and Error Status for top 10 status codes. Note that `FileNotFound' error and `ModuleNotFoundError' are specific to Python3.x. 

The overall execution success rate is 27.92\% considering the code snippets that ran in either one of the environments. 20.74\% of the snippets ran in both environments. Of the codes that ran in Python2, 19.59\% did not run in Python3 and of the codes that ran in Python3, 9.29\% did not run in Python2. This may be due to the fact that from Python2 to Python3 there has been syntax changes like for the print function. For example, both of the below commands are executable in Python2. However, only the latter is executable in Python3.

\begin{lstlisting}[language=Python]
print `Hello,World!'
print(`Hello,World!')
\end{lstlisting}


\begin{table}[]
\begin{tabular}{@{}lrrrr@{}}
\toprule
Status Name & \multicolumn{1}{l}{Py2 Count} & \multicolumn{1}{l}{\%} & \multicolumn{1}{l}{Py3 Count} & \multicolumn{1}{l}{\%} \\ \midrule

SyntaxError & 80451 & 29.82 & 97475 & 36.13 \\
NameError & 77671 & 28.79 & 76104 & 28.21 \\
Success & 69593 & 25.80 & 61689 & 22.87 \\
IndentationError & 14742 & 5.46 & 8228 & 3.05 \\
UnkownError & 11742 & 4.35 & 13279 & 4.92 \\
ImportError & 3619 & 1.34 & 507 & 0.19 \\
EOFError & 2604 & 0.97 & 1713 & 0.63 \\
FileNotFoundError & N/A & N/A & 4314 & 1.60 \\
TypeError & 1763 & 0.65 & 1699 & 0.63 \\
ModuleNotFoundError & N/A & N/A & 2751 & 1.02 \\ \bottomrule

\end{tabular}
\caption{Top 10 Status Code (Success and Error)}
\label{tab:executability-results}
\end{table}

\begin{table}[]
\centering
\begin{tabular}{@{}ccr@{}}
\toprule
\multicolumn{1}{l}{\textbf{Py2 Execution}} & \multicolumn{1}{l}{\textbf{Py3 Execution}} & \multicolumn{1}{l}{\textbf{Count}} \\ \midrule
Yes & Yes & 55960 \\
Yes & No & 13633 \\
No & Yes & 5729 \\
No & No & 194462 \\ \bottomrule
\end{tabular}
\caption{Comparison between Python2 and Python3 Executability}
\label{tab:truth-table}
\end{table}

\begin{figure}[htbp]
\centerline{\includegraphics[width=0.9\columnwidth]{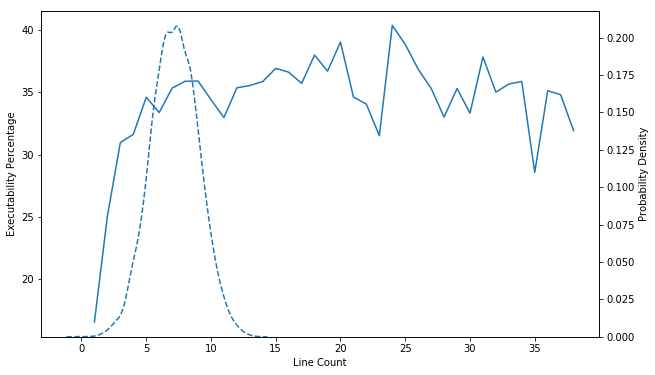}}
\caption{Change in executability with line count (solid line), line count distribution (dashed line).}
\label{fig:line-exec}
\end{figure}

We calculated the bootstrap difference between the executability of samples associated with accepted and non-accepted answers. We used 10000 iterations resulting in a 95\% CI of [-2.16,-1.36] and an average of $-1.75$ percent. This result doesn't show any significant difference between the executability of accepted and non-accepted answers. We also performed Wilcoxon test which gave us a p-value of 0.2163, showing the same result. We used R 3.5 for the statistical tests.

\Cref{fig:line-exec} depicts how executability changes with line count along with line count distribution. Snippets with line count of less than 3 has relatively lower executability. This may be due to the fact that these are often code outputs. 

\textit{RQ2: Can we find a trend which reflects how the executability of python snippets has changed over time and make conclusions from that?}

\begin{figure}[htbp]
\centerline{\includegraphics[width=0.9\columnwidth]{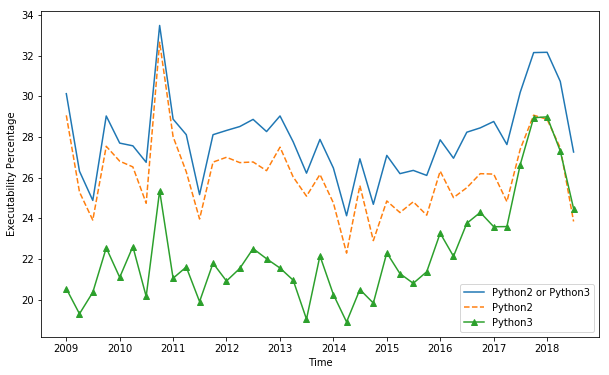}}
\caption{Change in executability over time.}
\label{fig:exec}
\end{figure}

\Cref{fig:exec} shows how Python2, Python3 and overall executability have changed in time. We fitted a linear regression model to the data and the coefficients show an increase of 0.1\% per year for overall, a decrease of 0.1\% per year for Python2 and an increase of 0.5\% per year for Python3 executability. This shows that the executability of the python code snippets doesn't change significantly with time.

\begin{figure}[htbp]
\centerline{\includegraphics[width=0.9\columnwidth]{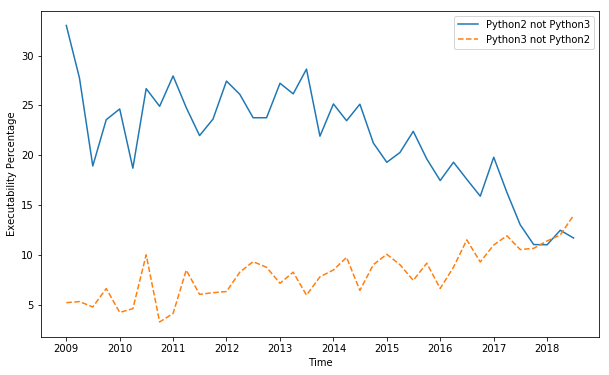}}
\caption{Cross platform change in executability over time.}
\label{fig:diff-exec}
\end{figure}

\Cref{fig:diff-exec} shows how the executability has changed for snippets that are executable in `Python2 but not in Python3' and executable in `Python3 but not in Python2'. Again, We fitted a linear regression model to the data and found that for snippets that are executable in `Python2 but not in Python3', there is a decrease of 0.42\% per year. On the other hand, for snippets that are executable in `Python3 but not in Python2', there is a increase of 0.21\% per year. This shows that Python3 is still being adopted by the stack overflow community.

\textit{RQ3: Can we find a relationship between executability of code snippets and GitHub reference?}

We found that the executability rate of snippets referenced in Github is 35.09\% which means that they are 25.86\% more likely to be executable than non-referenced ones.
We calculated the bootstrap difference between the executability of samples with and without GitHub reference. We used 10000 iterations resulting in a 95\% CI of [3.69,10.77] and an average of 7.2 percent. This shows us that code snippets with GitHub references tend to have a higher executability. We also performed Wilcoxon test which confirmed this result with a p-value of 5.985e-06.

\section{Discussion}
The results show that minor adjustments like proper module installation, can result in an executability of almost 28\%. An answer being accepted doesn't play a vital role in the snippet's chance of being executable, but a GitHub reference back to it does play an important role. This shows that executable code snippets are more likely to be used in GitHub projects. When it comes to temporal relationship with executability, for Python2 it is slightly decreasing while increasing for Python3. We believe this is due to the adoption of python3 among the stack overflow community. 

Regarding the threats to validity, we used answers corresponding to questions with only python tags, leaving out answers with composite tags. For example, a combination of python, matrix, numpy, scipy and scikit-learn tags are left out in this work. In addition, the top 40 module that we installed have the latest versions available at experimentation period (Feb-4-2019). Installing different versions of these libraries might change the results of executability obtained. 
Our results are biased towards the two specific versions of Python that we used.


\section{Conclusion}
We have developed a scalable docker based evaluator for checking the executability of python snippets. This can be used to perform such tests on a larger scale.
Our results show that executability doesn't affect the chances of the answer to be selected as the accepted answer significantly. This is probably due to the fact that the explanation is more important for an answer generally than being executable with little effort.
However, we observed that codes that have been referenced in a GitHub project showed a significantly higher executability rate. This is mainly due to the fact that for an answer to be used in a GitHub project, users tend to care more about them being executable with fewer changes required. For future work, we plan to expand our evaluator to include a larger dataset including composite tags and other languages using the same approach.

\section{Acknowledgement}
This research was enabled in part by support provided by WestGrid (www.westgrid.ca) and Compute Canada (www.computecanada.ca).

\balance
\bibliographystyle{ieeetr}

\end{document}